\documentclass[aps,prd,twocolumn,superscriptaddress,preprintnumbers,nofootinbib]{revtex4-1}

\usepackage{graphicx}
\usepackage{bm}
\usepackage{times}
\usepackage{slashed}
\usepackage{amssymb}
\usepackage{color}
\usepackage{slashed}
\usepackage{lipsum}
\usepackage{subfigure}
\usepackage{multirow}
\usepackage{amsmath}
\usepackage{array}
\usepackage{varwidth}
\usepackage{float}
\newcommand{\B}{\mathrm{\bf B}}

\begin{document}

\title{The Radar Echo Telescope for Cosmic Rays: Pathfinder Experiment for a Next-Generation Neutrino Observatory}
\author{S. Prohira}
 \email{prohira.1@osu.edu}
 \affiliation{Department of Physics, Center for Cosmology and AstroParticle Physics (CCAPP), The Ohio State University, Columbus OH 43210, USA}
  \author{K.D.~de~Vries}
\email{Krijn.de.Vries@vub.be}
  \affiliation{Vrije  Universiteit  Brussel, Dienst ELEM, IIHE,  Brussels,  Belgium}

\author{P.~Allison}
 \affiliation{Department of Physics, Center for Cosmology and AstroParticle Physics (CCAPP), The Ohio State University, Columbus OH 43210, USA}
\author{J.~Beatty}
 \affiliation{Department of Physics, Center for Cosmology and AstroParticle Physics (CCAPP), The Ohio State University, Columbus OH 43210, USA}
 \author{D.~Besson}
 \affiliation{University of Kansas, Lawrence, KS 66045, USA}
 \affiliation{National Research Nuclear University, Moscow Engineering Physics Institute, Moscow, Russia}
 \author{A.~Connolly}
  \affiliation{Department of Physics, Center for Cosmology and AstroParticle Physics (CCAPP), The Ohio State University, Columbus OH 43210, USA}
 \author{P.~Dasgupta}
  \affiliation{Universit\'{e} Libre de Bruxelles, Brussels, Belgium}
  \author{C.~Deaconu}
\affiliation{Enrico Fermi Institute, Kavli Institute for Cosmological Physics, Department of Physics, University of Chicago, Chicago, IL 60637, USA}

  \author{S.~De~Kockere}
  \affiliation{Vrije  Universiteit  Brussel, Dienst ELEM, IIHE,  Brussels,  Belgium}
  \author{D.~Frikken}
  \affiliation{Department of Physics, Center for Cosmology and AstroParticle Physics (CCAPP), The Ohio State University, Columbus OH 43210, USA}
  \author{C.~Hast}
 \affiliation{SLAC National Accelerator Laboratory, Menlo Park, CA 94025, USA}
 \author{E.~Huesca~Santiago}
 \affiliation{Vrije  Universiteit  Brussel, Dienst ELEM, IIHE,  Brussels,  Belgium}
 \author{C.-Y.~Kuo}
 \affiliation{National Taiwan University, Taipei, Taiwan}
 \author{U.A.~Latif}
 \affiliation{University of Kansas, Lawrence, KS 66045, USA}
   \affiliation{Vrije  Universiteit  Brussel, Dienst ELEM, IIHE,  Brussels,  Belgium}
  \author{V.~Lukic}
  \affiliation{Vrije  Universiteit  Brussel, Dienst ELEM, IIHE,  Brussels,  Belgium}
 \author{T.~Meures}
 \affiliation{University of Wisconsin-Madison, Madison, WI 53706, USA}
 \author{K.~Mulrey}
 \affiliation{Vrije  Universiteit  Brussel, Astrophysical Institute,  Brussels,  Belgium}
 \affiliation{Department of Astrophysics/IMAPP, Radboud University, P.O. Box 9010, 6500 GL Nijmegen, The Netherlands}
 \author{J.~Nam}
 \affiliation{National Taiwan University, Taipei, Taiwan}

\author{A.~Nozdrina}
\affiliation{University of Kansas, Lawrence, KS 66045, USA}
\author{E.~Oberla}
\affiliation{Enrico Fermi Institute, Kavli Institute for Cosmological Physics, Department of Physics, University of Chicago, Chicago, IL 60637, USA}
\author{J.P.~Ralston}
\affiliation{University of Kansas, Lawrence, KS 66045, USA}
 \author{C. Sbrocco}%

 \affiliation{Department of Physics, Center for Cosmology and AstroParticle Physics (CCAPP), The Ohio State University, Columbus OH 43210, USA}

\author{R.S.~Stanley}
 \affiliation{Vrije  Universiteit  Brussel, Dienst ELEM, IIHE,  Brussels,  Belgium}
\author{J.~Torres}
 \affiliation{Department of Physics, Center for Cosmology and AstroParticle Physics (CCAPP), The Ohio State University, Columbus OH 43210, USA}
 \author{S.~Toscano}
  \affiliation{Universit\'{e} Libre de Bruxelles, Brussels, Belgium}
  \author{D.~Van~den~Broeck}
  \affiliation{Vrije  Universiteit  Brussel, Dienst ELEM, IIHE,  Brussels,  Belgium}
   \affiliation{Vrije  Universiteit  Brussel, Astrophysical Institute,  Brussels,  Belgium}
  \author{N.~van~Eijndhoven}
  \affiliation{Vrije  Universiteit  Brussel, Dienst ELEM, IIHE,  Brussels,  Belgium}
 \author{S.~Wissel}
\affiliation{Departments of Physics and Astronomy \& Astrophysics, Institute for Gravitation and the Cosmos,   Pennsylvania State University, University Park, PA 16802, USA}
\affiliation{California Polytechnic State University, San Luis Obispo CA 93407, USA}
\collaboration{Radar Echo Telescope}

\begin{abstract}
The Radar Echo Telescope for Cosmic Rays (RET-CR) is a recently funded experiment designed to detect the englacial cascade of a cosmic-ray initiated air shower via in-ice radar, toward the goal of a full-scale, next-generation experiment to detect ultra high energy neutrinos in polar ice. For cosmic rays with a primary energy greater than 10\,PeV, roughly 10\% of an air-shower's energy reaches the surface of a high elevation ice-sheet ($\gtrsim$2\.km) concentrated into a radius of roughly 10\,cm. This penetrating shower core creates an in-ice cascade orders of magnitude more dense than the preceding in-air cascade. This dense cascade can be detected via the radar echo technique, where transmitted radio is reflected from the ionization deposit left in the wake of the cascade. RET-CR will test the radar echo method in nature, with the in-ice cascade of a cosmic-ray initiated air-shower serving as a test beam. We present the projected event rate and sensitivity based upon a three part simulation using CORSIKA, GEANT4, and RadioScatter. 
  RET-CR expects $\sim$1 radar echo event per day. 
\end{abstract}

\maketitle 

\section{Introduction}
Ultra high energy cosmic rays (UHECR) and neutrinos (UHEN) are important messengers from the most energetic astrophysical sources. Identifying and understanding these sources is a key goal of current multi-messenger astronomy, a burgeoning field with exciting recent breakthroughs and many discoveries to be made~\cite{mma, mma_ligo, blazar_icecube}.

The primary challenge to detecting UHECR and particularly UHEN is the low flux at the highest energies. This low flux requires an observatory that can efficiently probe a large target volume, in order to acquire a statistically significant sample of events. There are several current and proposed experimental strategies to achieve this large volume~\cite{rice, ara, anita,arianna,mountaintop,poemma,grand}. In this paper we discuss the {\it radar echo} method. This method has promising projected sensitivity to neutrinos in the 10-100\,PeV range, providing complementarity with existing and future techniques for measuring UHEN~\cite{chiba, chiba2,krijnkaelthomas, radioscatter,t576_run1, t576_run2}.
\begin{figure}[h]
  \centering
  \includegraphics[width=0.23\textwidth]{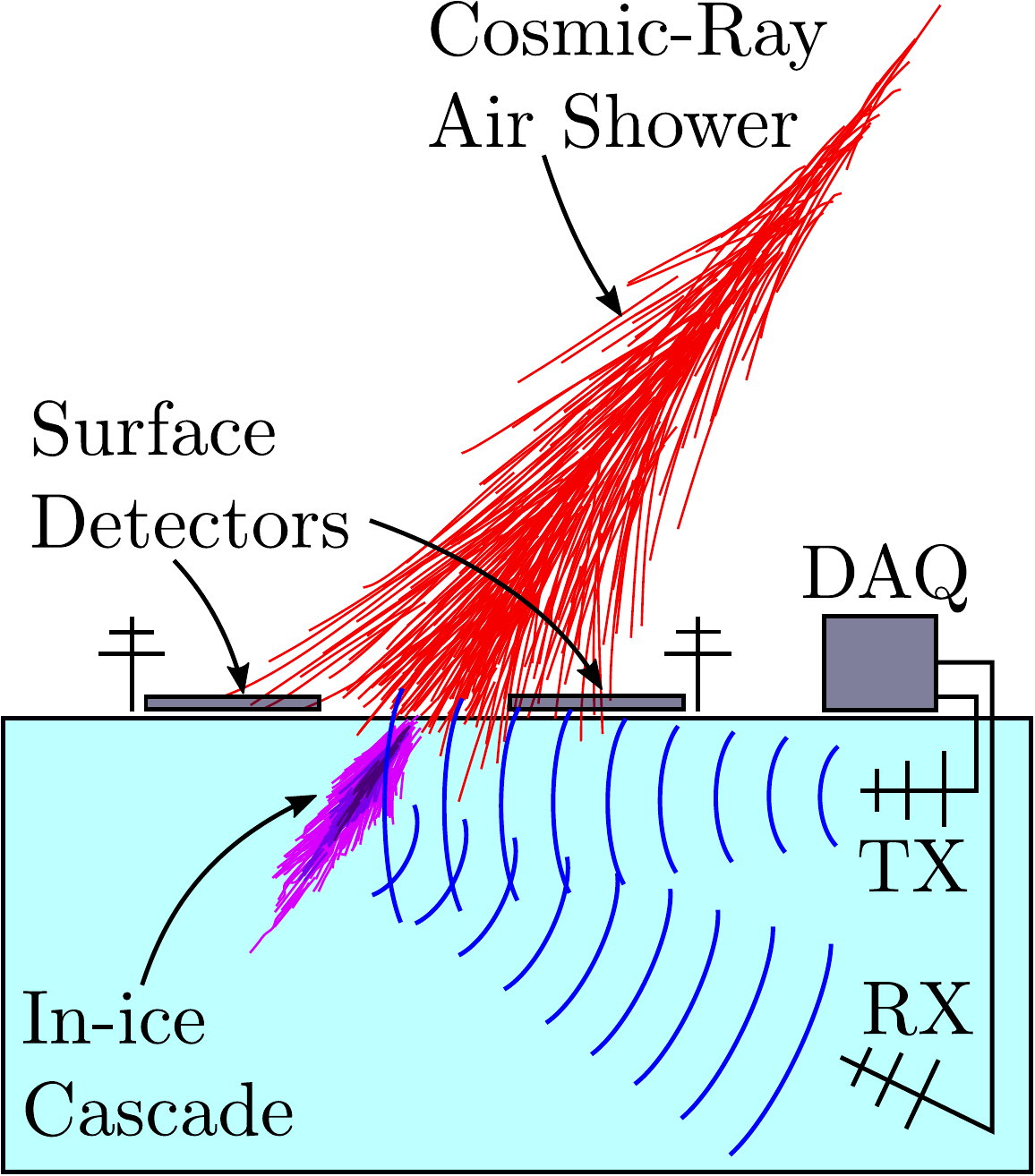}
    \caption{The RET-CR concept. A surface detector triggers on the charged in-air cascade particles as they reach the surface. The energy of this in-air cascade is deposited into the ice, where a denser in-ice cascade is produced. Radio is broadcast from the transmitter (TX) and reflected from the in-ice cascade to the embedded receiver (RX).}
  \label{concept}
  \vspace{-.6cm}
\end{figure}

A high energy particle (e.g. neutrino) interacting in a dense medium (e.g. ice) creates a cascade of relativistic particles that ionize atoms in the target medium. A short-lived cloud of charge is left behind, which can, if sufficiently dense, reflect incident radio waves. RET proposes to illuminate a volume of ice with a transmitter radio-frequency antenna and monitor that same volume for reflections with a receiving antenna. 
To improve reconstruction of the geometry of a cascade, and therefore the progenitor source direction, a target volume can be illuminated with multiple transmitters, and monitored with multiple receivers. 
Overall, the radar echo method allows for the coverage of a large volume with minimal apparatus and station layout optimized for a given neutrino energy, making it an attractive option for UHEN detection. 

The radar echo method has been verified in the lab \cite{chiba, chiba2}, with first observations of radar echoes from particle cascades recently reported\cite{t576_run1,t576_run2}. These lab tests are critical steps 
in developing an ultimate radar echo neutrino observatory. The final step in validating the technique is to translate the laboratory tests into nature, and test the method {\it in situ} with a known test beam: the in-ice cascade produced when the extensive air shower (EAS) of an UHECR impacts the ice. 

In this paper we describe the Radar Echo Telescope for Cosmic Rays (RET-CR) (Figure~\ref{concept}) which will serve as a testbed for the radar echo method, and a final stepping stone toward the eventual construction of a full-scale radar echo telescope for UHEN.

\section{History, theory, and background.}

First efforts on the radar echo method were chronicled by Lovell~\cite{early_radar}. With collaborator Blackett, 
the Jodrell bank observatory was constructed in the UK, anticipating that radar echoes from UHECR might explain ``sporadic radio reflexions'' from the upper atmosphere. Ultimately, those signals were determined to be from meteors, which ionize similar, far denser trails in the upper atmosphere. After several experimental efforts, including the Telescope Array RAdar (TARA) experiment, failed to detect UHECR via radar~\cite{tokyo_large_air_shower, mu_radar_1, tara, tara_limit}, and theoretical work explaining the lack of observed reflections~\cite{stasielak, bakunov}, the in-air method was finally deemed not viable due to short ionization lifetimes in the atmosphere at EAS altitudes, and damping from collisions between ionized electrons and neutral air molecules (an issue first raised by Eckersley in 1941~\cite{early_radar}, though largely subsequently ignored.)

Neutrino interactions in the ice produce ionization densities many orders of magnitude more dense than those in-air, owing to the $\sim 10^3$ greater density of ice relative to air. Therefore, while the ionization lifetime remains short in ice (roughly 10\,ns~\cite{ice_properties}) and the collision rate is extremely high, so too is the underlying ionization density, allowing for a possible scatter. Several models now exist for the in-ice radar echo~\cite{krijnkaelthomas, radioscatter} and show promising experimental sensitivity. Laboratory tests have shown good agreement with theory, but in order to test the radar echo method in nature, a source of {\it in-ice} ionization is required; EAS offer such a source.

\begin{figure}[t]
  \centering
  \includegraphics[width=0.48\textwidth]{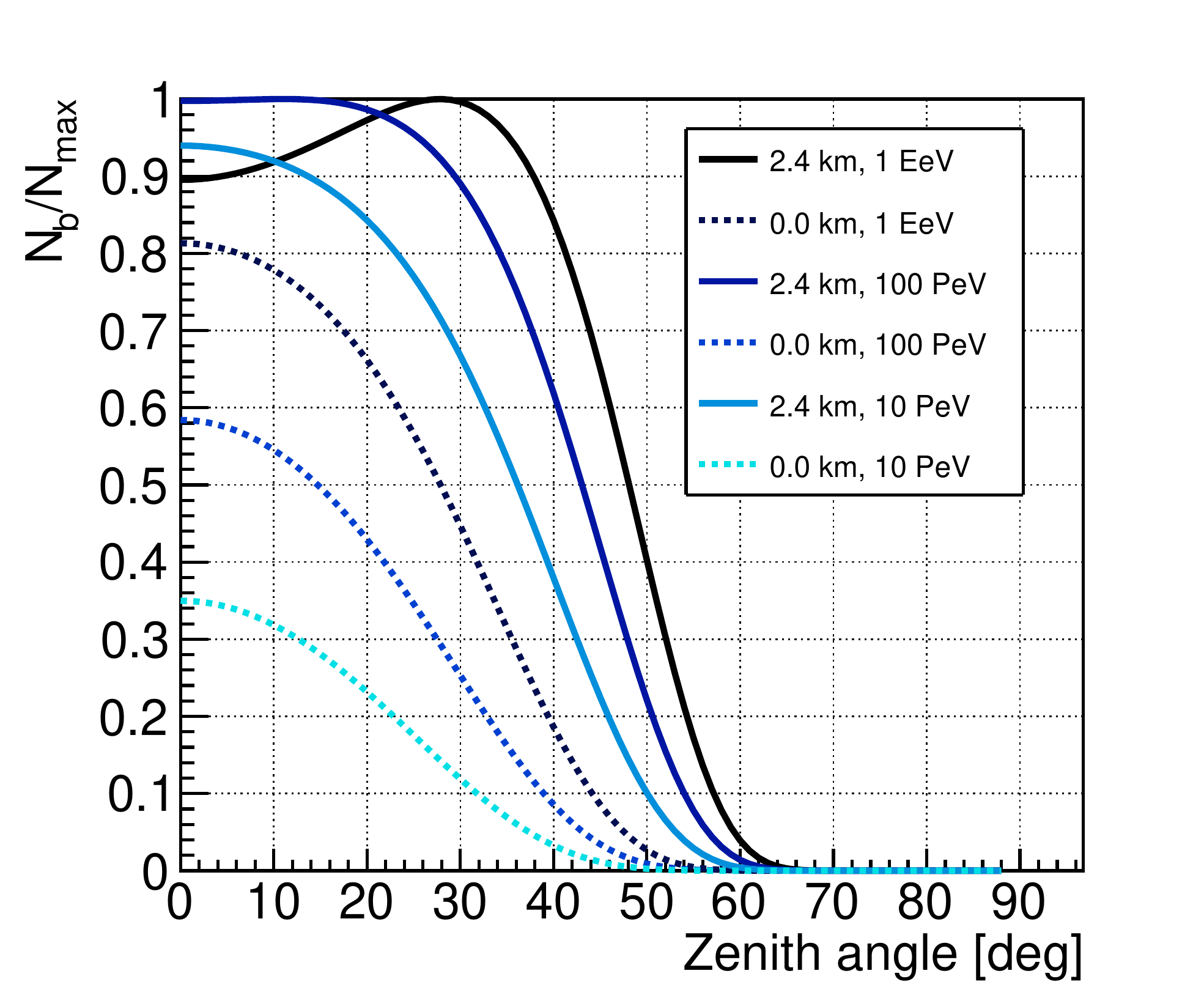}
    \caption{The ratio of the number of cascade particles, N$_b$, arriving at the air/ground boundary relative to the maximum number in the cascade, N$_{max}$, for various energies and ground elevations. At 2.4 km elevation and 1 EeV primary energy, shower maximum at normal incidence is in-ice.}
  \label{N_b_N_max}
\end{figure}

The EAS of a UHECR expands radially outward from the shower axis, such that an EeV cascade has a $\cal{O}$(100\,m) footprint on the ground. However, nearly all of the cascade energy is contained within $\sim$10\,cm of the shower axis, as illustrated below~\cite{DeKockere_in_ice_shower_core}.
The sought-after signal depends on the total number of particles in the shower core, as well as the geometry of the shower. 
In Figure~\ref{N_b_N_max}, we plot the ratio of the number of particles at the air/ground boundary (N$_b$) to the maximum number in the cascade (N$_{max}$) for different primary UHECR energies for different ice elevations as a function of zenith angle using the NKG approximation~\cite{kamata1958lateral, greisen1960cosmic}. For a high elevation (such as the interior Antarctic plateau), this ratio is $\geq$0.9 for energies $\geq$10\,PeV for a wide zenith angle range, indicating that a significant fraction of the energy of the primary particle will arrive at the air/ice boundary.

The core of a UHECR when it impacts the ice can be used as an in-nature test beam. Though TARA demonstrated that UHECR detection {\it in air} via radar echo was not feasible, the {\it in-ice} cascade produced by the remaining EAS particles,
as demonstrated by our beam tests, should be detectable via radar echo (for discussion of how the in-ice casade may be detected by Askaryan-type detectors, see Refs.~\cite{seckel_air_shower_core, krijn_eas}). For energies above 10\,PeV, we expect an ionized column with a density that decreases rapidly with radius, and an in-ice length of about 10\,m. The effective radius along this column at which radar will scatter depends on the transmitter frequency (discussed below); for frequencies in the 100s of MHz range, this radius is approximately 10\,cm. The profile of this ionization deposit is similar to that which would be produced by a neutrino-induced cascade, the primary difference being that neutrino events are more likely to occur in deep ice rather than near the surface.

The properties of the {\it in-ice} cascade from EAS have been studied using CORSIKA~\cite{corsika} to simulate the extensive air shower evolution, GEANT-4 \cite{geant} to simulate the propagation of these cascades once they enter the ice, and RadioScatter~\cite{radioscatter_github} to calculate the reflected signals from the ionization deposits left in the wake of the cascades. We next discuss our planned detector layout and design, and then describe our simulation and projected sensitivity.

\section{Experimental concept} The experimental concept is shown in Figure~\ref{concept}. A transmitter illuminates the region of ice just below the surface, with receivers monitoring this same region. The EAS of a UHECR with a primary energy greater than $10^{16}$\,eV deposits a fraction ($\gtrsim$10\%) of the primary energy at the surface of a high-elevation ice sheet. This energy is largely centered around the cascade axis, resulting in a dense secondary in-ice cascade roughly 10\,m long. The charged particles from the EAS are detected by a surface scintillator array, triggering a radio receiver waveform to be recorded in the radar data acquisition system (DAQ). This simple setup closely parallels that already employed for the laboratory test-beam experiment, with the focus of the experiment on post-run, offline analysis of the data. This relative simplicity also allows for testing various radar based trigger routines, which can be evaluated against the charged particle trigger. Such testing is critical, as an eventual neutrino detector will be triggered by the radar signal itself.

\begin{figure}[b]
  \centering
  \includegraphics[width=0.48\textwidth]{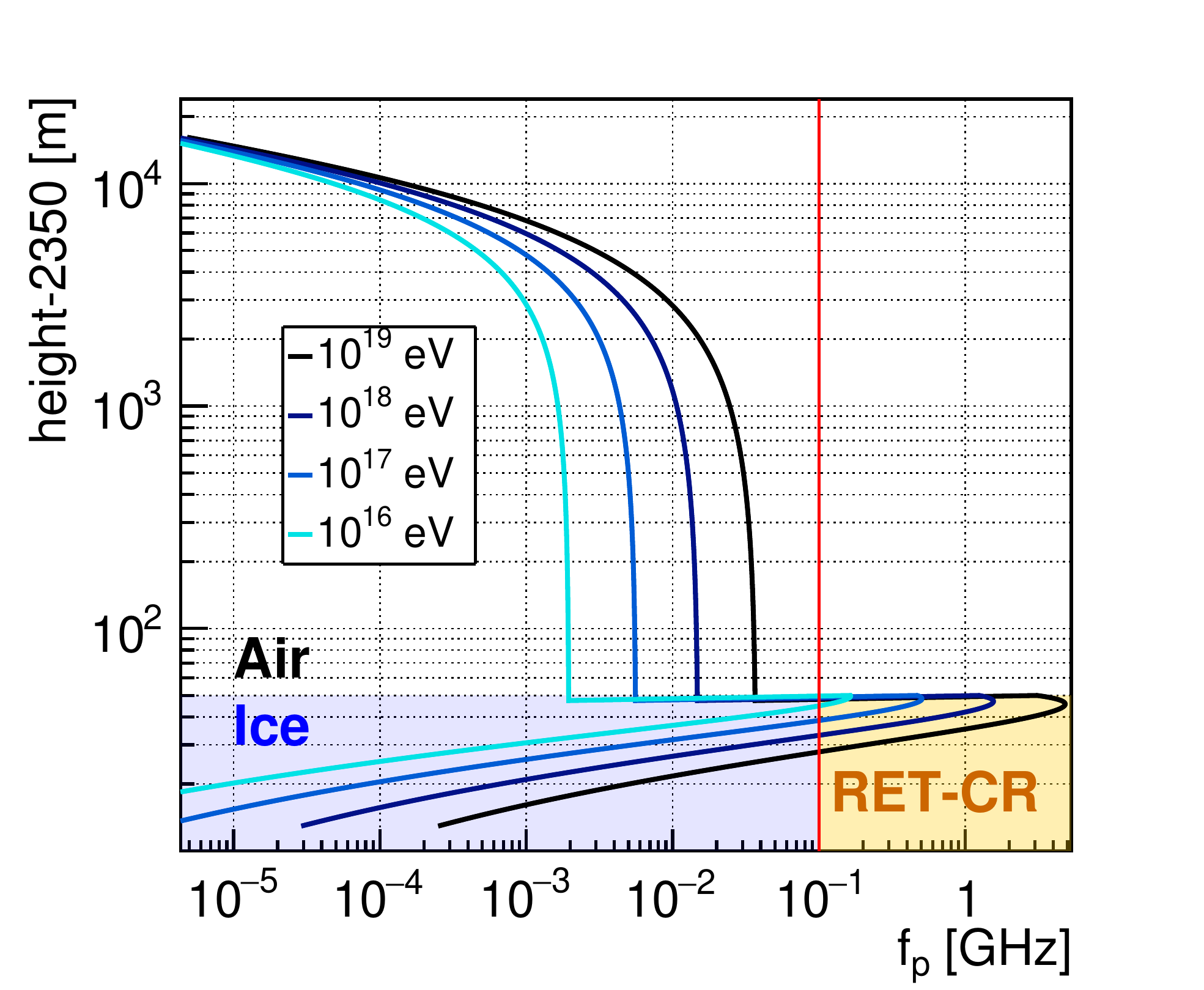}
    \caption{The peak plasma frequency ($f_p=\omega_p/2\pi$) of the cascade as it develops in-air and then in-ice, at normal incidence. The in-ice cascade is far more dense than the in-air cascade, making it detectable via radar for primary energies greater than $10^{16}$\,eV. The red line indicates the point at which $f_p >$100\,MHz.}
  \label{plasmafreq}
\end{figure}

When the cascade leaves the air and enters the ice, the density of the resultant ionization increases dramatically. This results in an ionization deposit in the ice with a plasma frequency $\omega_p=\sqrt{4\pi n_e q^2/m}$ far higher than any point in air. The plasma frequency, with $n_e$ the number density of the ionization, $q$ the electric charge, and $m$ the electron mass, is a measure of the density of an ionization deposit. To first order, incident fields with interrogating frequencies lower than $f_p=\omega_p/2\pi$ are reflected efficiently. \footnote{In a collisionless plasma, $\omega_p$ is a hard cutoff between the 'overdense' and 'underdense' regimes, which indicate fully opaque (reflective) or semi-transparent plasma, respectively. When taking collisions into account (as we do in our simulations), this boundary is smeared, but $\omega_p$ is still a useful discriminator for the underlying ionization density.}  The profile of $f_p$ from a primary cosmic ray as it moves through air into the ice is shown in Figure~\ref{plasmafreq}, where the in-air and in-ice components of an EAS are indicated, as well as a vertical line indicating 100\,MHz. For the in-ice portion of EAS with primary energies greater than $10^{16}$\,eV, $f_p$ $\gtrsim$ 100\,MHz, indicating efficient scattering for interrogating frequencies in this range.

In the following sections, we describe the various sub-systems of RET-CR. We provide a detail of the experimental layout in Appendix~\ref{appendix:station_layout}.

\subsection{Surface detector}\label{surface}

The RET-CR surface detector is designed to detect the air shower incident on the surface of the radar detector, providing both an external trigger to the radar DAQ, and an independent reconstruction of the air shower. Using the surface detector as a trigger for the radar system will ensure that an UHECR has entered the radar detector volume with sufficient energy to be detected through the radar echo technique. The independent reconstruction of shower parameters by the surface detector will provide values for the core position, energy, and arrival direction of the incident UHECR. These values will then be used to validate the reconstruction parameters obtained by the radar echo system.

The primary component of the surface detector is a scintillator plate array. The plates will be grouped in pairs following the design of the Cosmic-Ray Energy Cross-Calibration Array ~\cite{crossCalibrationArray, lofarEnergyScale}. The plates in each pair of scintillators will be separated by 20\,m. Ths scintillators will be accompanied by a butterfly radio antenna operating in the frequency band 30-300\,MHz, to form a station. Each station will have its own DAQ and power system. The combination of a radio and scintillator signal at each station will be beneficial in providing event reconstruction and energy estimates (more details below). The current deployment layout is shown in Appendix~\ref{appendix:station_layout}, where the stations are grouped into two sets of three stations, separated by the central radar system. Additionally, a system diagram is provided in Appendix~\ref{appendix:system_diagram}. The station layout has been optimized for trigger efficiency, discussed below.

The current prototyping and simulation development work utilizes the scintillators from the KASCADE experiment ~\cite{kascade}. The butterfly radio antennas have been donated by the CODALEMA/EXTASIS experiment~\cite{codalema_extasis, codalema_geomagnetic}. As such, for modelling the polyvinyl-toluene scintillator in GEANT4 we use a carbon:hydrogen ratio of 9:10~\cite{lora}. The panels deployed to the field will be similar in size and composition, and we do not expect any difference in performance from the panels simulated here.

The scintillator trigger threshold is tuned to maximize the radar echo detections. Simulations indicate that air showers with energy less than $10^{15.5}$\,eV are inadequate to produce an in-ice cascade detectable via radar echo (and the rate of such showers would overwhelm the DAQ storage and may cause interesting events at higher energies to be lost). Additionally, simulations of the air shower radio footprint show that radio reconstruction of the air shower is not possible for showers with an energy less than approximately $10^{16.0}$\,eV.
Therefore we target 100\% efficiency at $10^{16.5}$\,eV, with efficiency defined as the percentage of cosmic ray air showers traversing the instrumented area that trigger the surface detector. 
We aim for a trigger rate of order $10^5$ events per month, leading to approximately 300 surface triggered events a day. This is a manageable rate for both the surface and radar DAQ systems.

Simulation studies have been conducted to determine an appropriate triggering scheme for the surface detector. Events have been simulated with energies in the range $10^{15.0}$ eV to $10^{19.0}$ eV  and zenith angles in the range 0-30 degrees. We limit ourselves to this zenith range because in-ice energy deposition decreases dramatically beyond 30 degrees zenith. At higher energies, cascades at zenith angles $>$30 degrees will likely be detectable via radar, and will increase our event rate slightly relative to what we present here (e.g. for EeV cascades, it could increase the rate by up to a factor of 2). The simulations were made using the CORSIKA and CoREAS~\cite{Huege:2013vt} software for air shower simulation with a ground elevation set to 2400\,m, that of an optimal deployment site, Taylor Dome, Antarctica. A change in this altitude will affect the point within the shower development at which the air shower passes through the detector. Showers at sea level are generally developed beyond shower maximum before reaching the ground. At the altitude of Taylor Dome, whether the shower has developed to a point before or after the shower maximum is strongly dependent on the energy and zenith angle of the air shower (Figure~\ref{N_b_N_max}).

\begin{figure}[h]
  \centering
  \includegraphics[width=0.48\textwidth]{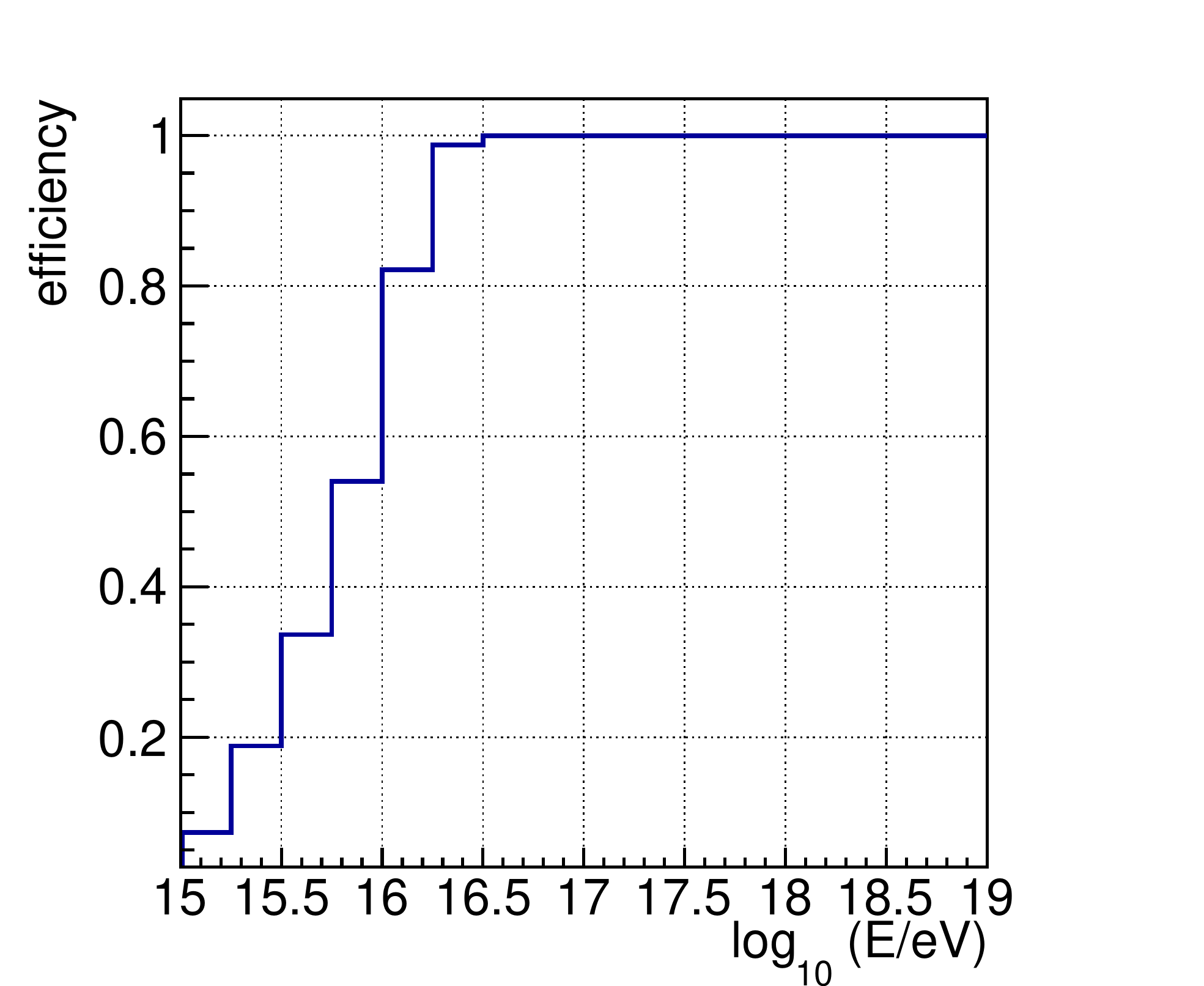}
    \caption{Trigger efficiency curve for the surface system as a function of primary UHECR energy.}
  \label{trigEffSurface}
\end{figure}

The stations of the surface detector will trigger independently. Each scintillator will be required to contain a deposit of 6 MeV (1 minimally ionizing particle) or greater per event and both scintillators in a station must trigger coincidentally within an event (an L0 trigger). 
The final trigger requirement is that all stations within one cluster must have coincident, above threshold energy deposits in all scintillators (an L1 trigger). The width of the time window for this coincidence, $\sim$170\,ns, corresponds to the maximum time-difference-of-arrival between two stations separated by roughly 100\,m for a $30^\circ$ zenith angle cascade. The resulting trigger efficiency is shown in Figure~\ref{trigEffSurface}. In this figure, we show that we achieve 100\% efficiency at $10^{16.5}$ eV, as desired. Decreasing a half-decade in energy, at $10^{16.0}$ eV the efficiency decreases to $\sim$70\%. In the lowest energy bin simulated, the efficiency is approximately 5\%. This rapid turn-on in our trigger threshold allows us to target our desired event rate. 

The radio component of the surface system operates at 1\,GS/s, allowing for precise angular reconstruction of cascades, which can then be used to test the reconstruction capabilities of the radar detector. Similar radio-based cosmic ray detectors with comparable baselines can reconstruct arrival direction with an error less than 1-2 degrees~\cite{lofarShape, lopesFinalResults, codalemaCurrentStatus, augerNanosecond}. At the time of writing, a set of three surface stations are taking data in a rooftop test configuration with $\sim$100\,m baselines. Analysis of these data will provide an accurate number on angular reconstruction of the surface system in advance of RET-CR.

Furthermore, the radio component of the surface system also provides a measurement of the energy of the cascade. Similar radio based cosmic ray detection experiments~\cite{tunkaRexReco, lofarEnergyScale,lopesFinalResults} including those performing reconstruction with a limited number of antennas~\cite{aeraEnergyReco} can constrain energy to approximately 15-20\%, with some studies~\cite{ariannaSingleStationReco} claiming 15\% resolution with just a single station. The RET-CR surface stations will have a slightly wider bandwidth than many of these experiments, so we expect a similar---if not slightly better---energy uncertainty. Our rooftop test data will provide an accurate number for the energy uncertainty of our specific system in advance of RET-CR.

To estimate the core reconstruction accuracy, we developed a reconstruction procedure using the realistic particle deposit simulated with CORISIKA.  For a given simulation, different core positions where chosen, and the scintillator deposit was determined, conservatively assuming a 15\% uncertainty in measured deposit.  We then used a minimization technique based on the Nishimura-Kamata-Greisen (NKG) function to reconstruct the core position, similar to what is done in~\cite{kascade}.  Using this method, we were able to reconstruct the core with a mode (68\% quantile) resolution of 7.8 (24.4)\,m at $10^16$\,eV, and 10.8 (24.6)\,m at $10^17$\,eV for showers with a zenith angle less than 30 degrees.  This is an upper bound on the core resolution, as we will develop more advanced core fitting methods making use of radio measurements, which should improve reconstruction for the outlier events that inflate the size of the 68\% interval.

\subsection{Radar echo detector}

\subsubsection{Data Acquisition System}

The primary element of the radar echo detector DAQ is a Xilinx RFSoC~\cite{rfsoc}. This all-in-one 2device will be used for both the transmitter and receiver components. The transmit portion comprises 8 channels, each with a 6\,GS/s digital-to-analog converter (DAC) capable of producing a phased, modulated output to an array of transmitters. The receive portion has 8 channels, each channel with a 4\,GS/s analog-to-digital converter (ADC). A Virtex-7 FPGA provides transmit and receive functions and trigger logic, and an on-board ARM processor facilitates information transfer between the FPGA and the communication subsystems, also described below. 

The DAQ will have local storage for buffering data and a prioritizer system for telemetry, with a design expectation that our data transfer rate will be the primary bottleneck in getting data from the station. The most promising events are sent via communication link as well as being stored locally on disk.
The criteria for this can include i) measured primary energy from the surface detectors or ii) the proximity of the in-ice vertex to a receive antenna, or a combination of the two.  The disks will be retrieved at the end of the season.

The FPGA will host a number of radar ``triggers'' with potentially varying topologies, even though these triggers do not actually signal an event snapshot. These triggers will be trained against the surface detector trigger to determine their efficacy for eventual use in the successor neutrino detector, where the radar signal itself must trigger the DAQ. One such trigger under investigation is a heterodyne trigger (also called a ``chirp trigger'') based upon a method developed for the TARA remote stations~\cite{remote_stations,firmware_trig} that exploits the frequency shift of the return signal. The geometry of RET-CR is such that all of the received radar echoes will exhibit this frequency shifting behavior. Other triggers based on the unique radar signature are also being explored. The sensitivity studies in this article employ a simple threshold trigger for the radar component.

\subsubsection{Transmit array and transmitter modulation}

The transmitter for RET-CR will be a vertical phased array of 8 vertically polarized antennas buried 2-20\,m below the ice surface. The exact depth requires further study of ice properties and a better understanding of radio propagation near the surface of the ice through ongoing simulations, as discussed in section~\ref{sec:radioscatterinice}. This phased array serves 2 critical functions. The first is directionality---a phased array governed by an FPGA can form high gain beams in a defined direction, achieved via adjusting the relative phases of the transmit signal being delivered to each of the antennas in the array. A vertical phased array has azimuthal symmetry with a high gain beam at a defined zenith angle, defined by the relative phase delays of each antenna. Since our reflectors are confined to the top $\sim$10\,m of the ice just below the surface, we can steer the beam slightly upward virtually no power is beamed to the region below, where we do not expect to receive UHECR core reflections. Recent studies in Antarctica have shown that in-ice phased arrays are highly efficient receivers~\cite{phased_array_trigger}, and phased transmitter arrays are common in use throughout the world, including the TARA experiment. The second critical function of a phased array is to lower the single-amplifier gain for the transmit power amplifier. In lieu of a single 160\,W power amplifier, each antenna will have its own 20\,W power amplifier. This distributes the ohmic heating losses over 8 antennas instead of 1 and provides some redundancy: in the event that a power amplifier malfunctions, the experiment loses some efficiency, but does not shut down entirely.

The antennas will be based upon the simple bicones or biconical dipoles used by the RICE, ARA, and ARIANNA experiments in Antarctica.
These antennas are small enough to fit down a borehole but are sufficiently broadband as to allow for a range of transmit frequencies and modulations. Simulations using FDTD~\cite{meep} and parabolic equation codes~\cite{paraProp, paraprop_github} are underway to determine if non-uniform antenna spacing, or antennas with asymmetric zenith angle gain can increase transmitter efficiency in the direction the beam is `steered'. Some recent studies also indicate that broadband phased arrays may be possible in ice~\cite{hanson_phased}, though focus here is on higher frequencies than those of interest to RET-CR.

The modulation scheme is currently being defined. We plan to frequency hop or frequency shift around a central carrier of 100-300\,MHz, with a transmitter bandwidth of 50-100\,MHz. 
The central frequency is determined by maximizing the signal to noise ratio of a radar signal to the background noise.  The signal has an optimal frequency dependent on the cascade dimensions and density, and the noise decreases with increasing frequency as thermal noise begins to dominate over galactic noise above $\sim$150\,MHz.

This central frequency and the ultimate modulation strategy will be determined via simulations that are already underway. Modulation, as opposed to pulsing, increases detector livetime, as long as the carrier signal can be removed from the receivers. We discuss this below in section~\ref{filtration}.

\subsubsection{Receive array}
The receiver array will be laid out in the configuration shown in Appendix~\ref{appendix:station_layout}. Two different TX-RX baselines allow for a wide range of primary particle energies to be detected. Similar to the transmitter array, the receiver antennas will be buried 2-20\,m below the surface of the ice. Each receive antenna will be a vertical phased array, similar to the transmit array, in order to maximize near-surface gain with full azimuthal coverage.

The receivers will not trigger the DAQ, but will form triggers as a testbed for eventual use in a neutrino array.
\subsubsection{Amplification and adaptive filtration}\label{filtration}

We will have a limiter and high-gain, low-noise amplifier on each receive channel, providing protection during transmitter turn-on and approximately 70\,dB of gain, respectively. This amount of gain is sufficient to attain the galactic noise floor at our frequency, location, and receiver bandwidth of $\sim$100\,MHz.

Because radar receivers will be illuminated by the transmitter, it is essential to filter the transmitter or gate the receivers such that amplifier saturation does not occur. We plan to adopt an adaptive filtration scheme, whereby we will record an amplifier-bypassed snapshot of the transmitted signal over a horizon-distance window in time at each receiver, and then inject it time-delayed and phase-inverted into the receiver stream before the amplifier chain. The delay and output amplitude are tuneable, allowing for an iterative reduction of the input amplitude until the carrier is fully eliminated. This procedure will be updated at intervals throughout the day to account for environmental changes such as snow accumulation, which have been shown to introduce measurable changes in reflection times on $\sim$day timescales~\cite{ariannaRecoSnow}.

\subsubsection{Power, system health, calibration, and communications}

The detector will be fully autonomous and powered by three 1.2\,kW solar arrays arranged in a triangle, such that at any time of (a sunny) day the station is provided with approximately 1\,kW of power, with the majority of this power being used by the transmit power amplifiers. A bank of batteries will buffer power to assist in running the stations during adverse weather conditions. Each surface station will be powered by an individual photovoltaic. RET-CR will run only during the austral summer. 

The system health, including power consumption, DAQ enclosure temperature, power amplification health, and local weather will all be monitored remotely, in real-time.

We will deploy a small, autonomous calibration unit that sends out a broadband pulse at regular intervals. This unit will serve as a regular baseline for thresholds and ensure global time synchronization, as well as for active monitoring of the above mentioned environmental changes, such as snow accumulation. 

The communication system will be a 2-way satellite-based internet link. Through this link we will telemeter the prioritized data and system health information back to the lab, and, from the Northern Hemisphere, new trigger schemes and other station software and commanding to RET-CR. Alternative communications links via point-to-point Ethernet may be possible if there are line-of-sight repeater stations between a major base and the remote RET-CR.

\begin{figure}[h]
  \centering
  \includegraphics[width=0.44\textwidth]{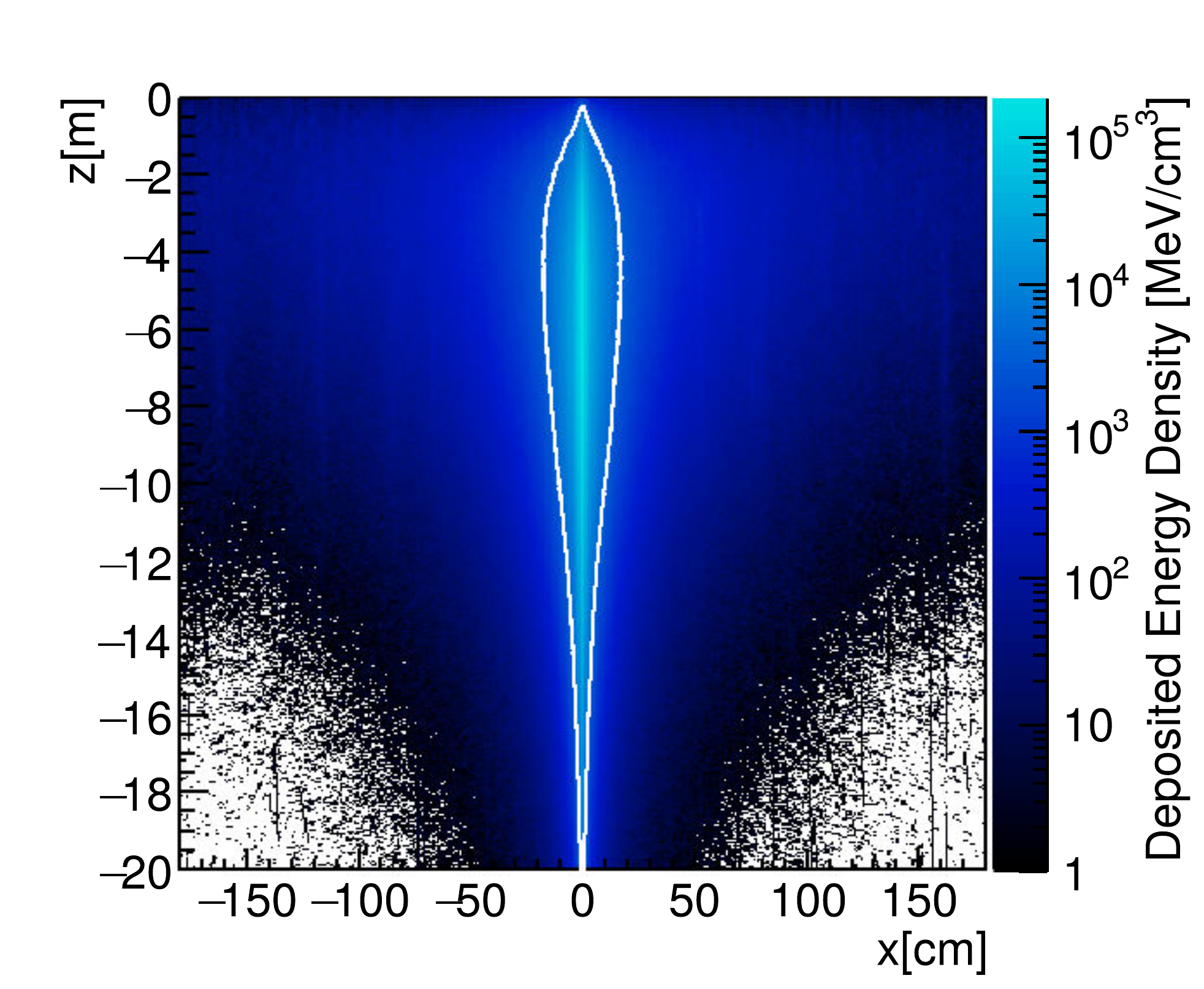}
    \caption{A one centimeter wide, two-dimensional slice of the in-ice energy density distribution along the cascade axis for a primary proton with E$=10^{17}$\,eV. The solid white line outlines the region for which the plasma frequency exceeds 100\,MHz.} 
  \label{exCascade}
\end{figure}

\section{Projected sensitivity}

\begin{figure*}[ht]
  \centering
  \includegraphics[width=0.98\textwidth]{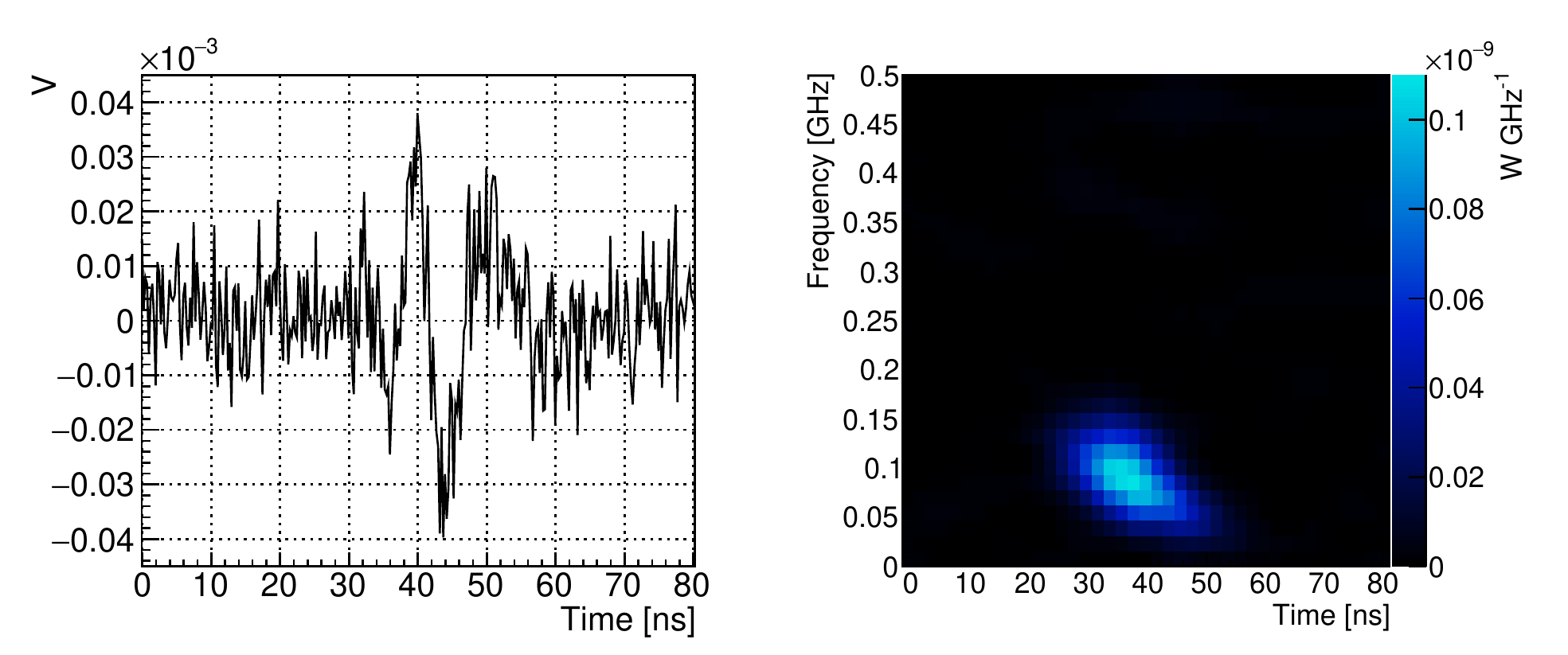}
    \caption{An example event simulated using CORSIKA+GEANT4+RadioScatter, as described in the text. This is a $10^{16.5}$\,eV primary at normal incidence with a 10\,ns plasma lifetime, simulated with a 160\,W transmitter at 100\,MHz.}
  \label{exampleEvent}
\end{figure*}

The approximate sensitive surface area instrumented by RET-CR is $5 \times 10^4$\,m$^2$. Through this area we can expect a flux of roughly 1 event at 100\,PeV per day. The surface system will trigger on every cosmic ray with a primary energy above this, with decreasing efficiency at lower energies, as described in section~\ref{surface}. To simulate our radar echo detection efficiency, we performed a detailed multi-step Monte Carlo which we describe here using (in order) CORSIKA, GEANT4, and RadioScatter.

\subsection{CORSIKA cascades and the surface stations}

A CORSIKA-based Monte Carlo simulation for optimizing the surface array location was described in section~\ref{surface}. This same distribution of events (core positions, zenith angles, and energies) was used to simulate the radar sensitivity of RET-CR. A separate set of CORSIKA simulations was prepared specifically for producing the GEANT4 output used in subsequent simulation steps. These CORSIKA showers were produced at 0, 15, and 30 degrees zenith for each half-decade energy, with a ground elevation of 2400\,m, as before. For $10^{16}$\,eV and $10^{16.5}$\,eV, no thinning was employed. For higher energies, thinning is set to $10^{-7}$ of the primary particle energy. For $10^{17}$ eV and $10^{17.5}$ eV the CORSIKA `weight' of a single particle will never be larger than 10, for $10^{18}$ eV it will never be larger than 100. Thinning retains the overall energy of a cascade, such that the total in-ice ionization number will be the same for any thinning, but it changes the distribution of low-energy particles in the final footprint (which is then used as the input to GEANT4). We therefore minimized thinning as much as possible, subject to computing constraints.

\subsection{GEANT4 simulations from CORSIKA output}

The CORSIKA particle output at the surface of the ice was subsequently used as input for the GEANT4 simulation code configured to propagate particles into the ice. For this a realistic density profile similar to that found at South Pole was used, $\rho(z) = 0.460 + 0.468 \cdot (1 - e^{-0.02z})$ with $\rho$ the density in g/cm$^3$ and $z$ the depth in m. In each step of the simulation, the ionization energy loss is recorded; taking a typical ionization energy of $50$~eV allowed us to obtain the free charge density profile in the ice~\cite{PDG2020}. An example of this profile is shown in Figure~\ref{exCascade} for an air shower induced by a $10^{17}$~eV proton primary incident on an ice sheet at 2400\,m elevation. 
The plasma frequency is a good indicator of the reflective properties of the induced plasma, as discussed previously. The solid white line in Figure~\ref{exCascade} outlines the region for which the plasma frequency has a value larger than 100\,MHz, where fully coherent scattering is expected.

\subsection{RadioScatter simulations of GEANT4 output}\label{sec:radioscatterinice}

To simulate the overall sensitivity, we use the RadioScatter~\cite{radioscatter_github} code. RadioScatter is a particle-level c++ code to simulate radio scattering from ionization deposits. It calculates the received radio signal reflected from an ionization deposit (from e.g. a particle cascade) for an arbitrary geometry of transmitter(s) and receiver(s).

The energy deposition calculated by GEANT4 was used as the input to RadioScatter. An example of a triggered event from a $10^{16.5}$\,eV primary at normal incidence is shown in Figure~\ref{exampleEvent}. Clearly visible is the characteristic frequency shift expected for the RET-CR geometry, which can be exploited in trigger routines. At each cascade position in the surface scintillator simulation set, we simulated two different GEANT4 cascades: 1) the cascade with the closest half-decade energy {\it below} the primary energy of the surface simulation and 2) the cascade with the closest half-decade energy {\it above} the primary energy of the surface simulation. This is done because generating the GEANT4 cascades from the CORSIKA output is computationally expensive, so each discrete energy and zenith angle cannot be simulated individually. This method bounds the amount of energy that could arrive at the surface and accounts for shower-to-shower fluctuations. Both energies were simulated at each cascade position at one of three zenith angles, 0, 15, and 30 degrees selected according to their proximity to the `true' zenith angle of the cascade from the surface simulation. These cascades were generated with a uniform distribution in $\cos\theta$ because cosmic rays arrive isotropically at earth, and a uniform distribution in $\cos\theta$ ensures that any zenith angle dependencies of the trigger are reflected in the sensitivity.
The horizon distance for an in-ice transmitter is finite owing to the changing index of refraction in the firn (the top $\sim$100\,m of an ice sheet where snow is being compacted into ice)~\cite{kuivinen1983237,hawley2008rapid,ArthernGPR, densityToN}. We therefore put a hard cut on a horizon distance of 150\,m, which is commensurate with the point at which the in-ice shower maximum is out of view for a transmitter depth of $\sim$20\,m. The simulations in this paper eschew the typical ray-tracing approach for studying propagation in the firn since this has recently been shown to be incomplete without in-situ studies of the ice density profile~\cite{horizontal_propagation_FDTD,paraProp}. 
The hard horizon cutoff for the results presented here are a proxy for the loss in efficiency due to propagation effects. These effects will be explored in detail in a future work.

\subsection{Calculation of the event rate for RET-CR}

The two components of the event rate are the effective area of the detector and the cosmic ray flux. We define both of these over an energy bin $E_i$ with index $i$ and width $dE$. The effective area $\mathcal{A}_i^{eff}(E_i)$[m$^2$] for energy bin $i$ is a function of the cross sectional area $\mathcal{A_{\perp}}$ over which the sensitivity is calculated and a dimensionless trigger efficiency for the same bin, $\mathcal{T}_i(E_i)$. 
\begin{figure*}[ht]
  \centering
  \includegraphics[width=0.98\textwidth]{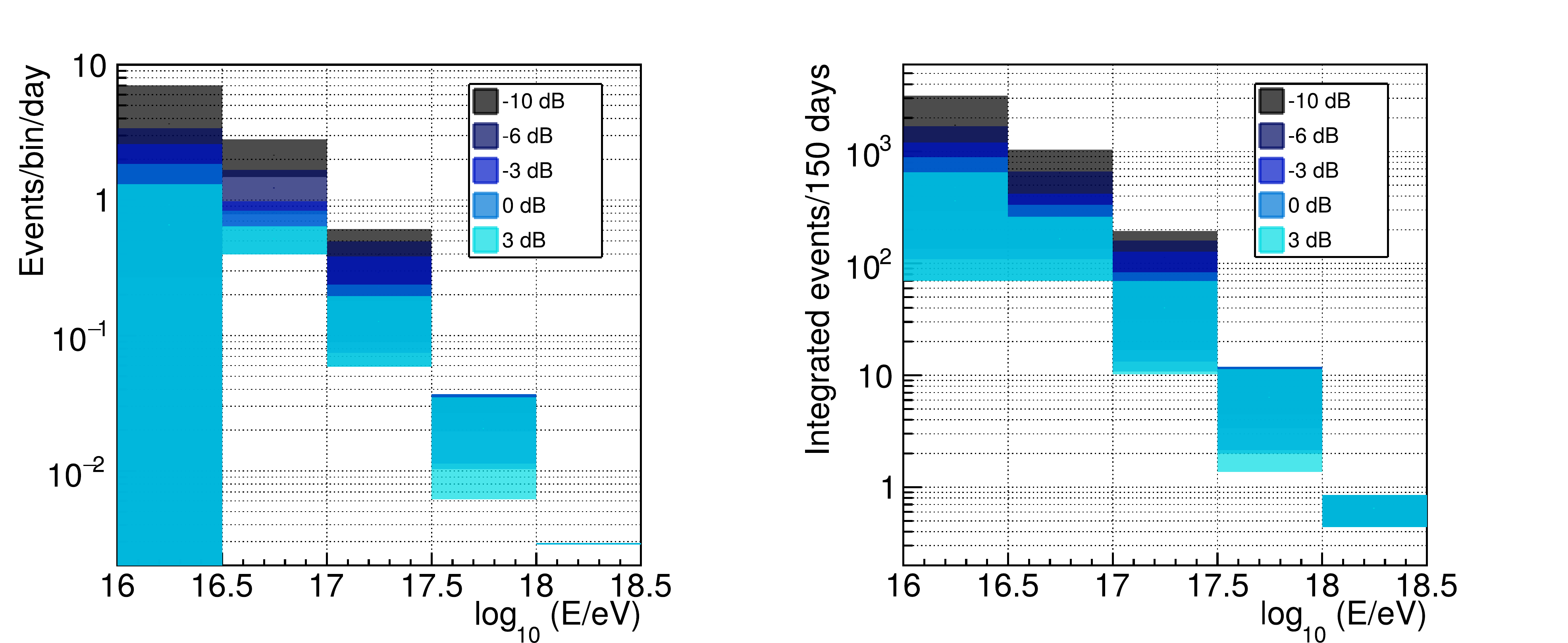}
    \caption{(Left) Event rates per day as a function of energy for the RET-CR detector. Colors correspond to different signal-to-noise relative to thermal noise of 8$\mu$V for a 100\,MHz bandwidth. Width of the bands are explained in the text. (Right) Integrated event rates per 150 days as a function of energy for the RET-CR detector. Each bin represents the total integrated number of events per 150 days at and above that energy. Colors correspond to different signal-to-noise relative to thermal noise. }
  \label{rates}
\end{figure*}
To detail the effective area we first introduce the boxcar function, $\B$, which bounds a number $x_1<x<x_2$:
\begin{equation}
\B(x, x_1, x_2)=\Theta(x-x_1)\Theta(x_2-x),
\end{equation}
where $\Theta$ is the Heaviside step function. Then we define our trigger conditions $\delta^{\mathrm{\bf S}}$ (for surface) and $\delta^{\mathrm{\bf R}}$ (for radar) which are 1 if the trigger condition is satisfied, and 0 if not. For example,
\begin{equation}
  \delta^{\mathrm{\bf R}}=\Theta(v^{peak}-v^{thresh})
\end{equation}
for a peak waveform voltage $v^{peak}$ and threshold voltage $v^{thresh}$, and $\delta^{\mathrm{\bf S}}$ is 1 when the surface system trigger logic is satisfied (a coincidence between surface stations with a certain per-station energy threshold, as described in Section~\ref{surface}). We then define the number of detected events, for event index $k$, energy bin with index $i$, and $\theta$ bin with index $j$, as a matrix $n_{ij}$. These events are weighted by $\cos\theta_k$ to account for the correction to the perpendicular cross-sectional area $\mathcal{A}_{\perp}$ seen by a cosmic ray at zenith angles greater than zero. 

\begin{multline}
      n_{ij}(E_i, \theta_j)=\\
      \sum_k\delta^{\mathrm{\bf S}}_k\delta^{\mathrm{\bf R}}_k\cos(\theta_k )\B(E_k, E_i, E_{i+1})\B(\theta_k, \theta_j, \theta_{j+1})
\end{multline}
$n_{ij}$ will be zero for bins in zenith outside of the aperture of the instrument. For RET-CR, this aperture is 0-30 degrees as discussed in section~\ref{surface}. 
The total number of simulated events $N_{ij}$, also as a matrix in $E$ and $\theta$ is
\begin{equation}
  N_{ij}(E_i, \theta_j)=\sum_k\B(E_k, E_i, E_{i+1})\B(\theta_k, \theta_j, \theta_{j+1}).
\end{equation}
The trigger efficiency $\mathcal{T}_{ij}$ in energy and zenith bins is represented as the ratio of these two, 
\begin{equation}
  \mathcal{T}_{ij}(E_i, \theta_j)=\frac{n_{ij}}{ N_{ij}}
\end{equation}
and we can then sum over all $\theta$ to get this expression as a function of energy only,

\begin{equation}
  \mathcal{T}_i(E_i)=\sum_j \frac{n_{ij}}{ N_{ij}},
\end{equation}
meaning that the effective area for energy bin $E_i$ is
\begin{equation}
  \label{eq:aeff}
  \mathcal{A}^{eff}_i(E_i) = \mathcal{T}_i \mathcal{A}_{\perp}.
\end{equation}

The flux as a function of energy $\mathcal{F}(E)[m^{-2} s^{-1} sr^{-1} eV^{-1}]$ is a broken power-law fit to the measured CR flux by many experiments~\cite{PDG2020, iceTopCRSpectrum, augerCRSpectrum}.  To get a number of events per square meter, per second, per steradian, in energy bin $E_i$, we integrate $\mathcal{F}(E)$ over the energy range of bin $i$,

\begin{equation}
  \mathcal{F}_i(E_i)=\int_{E_i}^{E_{i+1}} \mathcal{F}(E) dE.
\end{equation}

Finally, the expression we use to calculate the event rate as a function of energy, for energy bin index $i$, $\mathcal{R}_i(E_i)$, is given in Eq.~\ref{eq:rate},
\begin{equation}
  \label{eq:rate}
  \mathcal{R}_i(E_i) = \mathcal{A}_i^{eff}\mathcal{F}_i \int dt \int d\Omega,
\end{equation}
where the integral over time is the detector live time, and $\int d\Omega=\int d\phi d(\cos\theta)=\int d\phi \sin\theta d\theta$ is the integral over the aperture of the instrument. For an experiment sensitive to cosmic rays from the full sky, $\int d\Omega=2\pi$\,sr; the aperture for RET-CR is from 0-30 degrees zenith, $\int d\Omega\approx0.26\pi$\,sr.

We note that the measured flux of cosmic rays differs per experiment. A global study seeking to quantify this fluctuation between experiments~\cite{dataDrivenCRFlux} shows roughly 20\% spread in measured energies between various experiments in our range of interest, leading to an uncertainty in the true flux. Therefore, to account for this uncertainty, we use the cosmic ray flux normalization in line with the mean of the global fit. A goal of the forthcoming Cross-Calibration Array~\cite{crossCalibrationArray} is to mitigate this uncertainty between experiments.

Figure~\ref{rates}, left, presents our expected event rate per day as a function of energy for various signal-to-noise levels relative to a thermal noise RMS of 8\,$\mu$V. This is for a 160\,W transmitter at 100\,MHz with a 10\,ns plasma lifetime, a likely plasma lifetime for polar ice near the surface~\cite{ice_properties}. The upper and lower bounds of the bands correspond to the over and underestimated energy simulation respectively. The mean of the 0\,dB SNR curve integrates to roughly 1 event per day. 

Figure~\ref{rates}, right, shows the integrated event rate for one austral running season, approximately 150 days. An entry here represents the integrated number of events detected per 150 days at and above that energy, at the indicated SNR. We expect e.g. $\sim$50 events at and above $10^{17}$\,eV per season at the 0\,dB threshold level.

For comparison to RET-N, the in-ice cascade energy for a $10^{17}$\,eV primary detected by RET-CR is roughly $10^{16}$\,eV. This cascade energy corresponds to that of a charged-current neutrino-nucleon interaction of $10^{16}$\,eV, or a neutral current neutrino-nucleon interaction at $\sim5\times10^{16}$\,eV for inelasticity $y\sim 0.2$. Thus, the primary cosmic ray energies probed with RET-CR are similar to those of neutrinos to be targeted with RET-N.

\section{Conclusion}

We have presented the Radar Echo Telescope for Cosmic Rays, a pathfinder {\it in-situ} detector to test the radar echo method. Using the dense in-ice shower core of a cosmic ray air shower as a test beam, RET-CR will train trigger routines, energy and direction reconstruction methods, and analysis techniques to be employed by an eventual full-scale next-generation neutrino detector. 
\vspace{-.5cm}
\section*{Software}
CORSIKA version 7.7400 (with QGSJETII-04 and URQMD 1.3cr), CORSIKA 7.7100 (with QGSJETII-04 and GHEISHA 2002d),  CoREAS version 1.4 with a typical Taylor Dome, Antarctica atmosphere, GEANT4 versions 10.5 and 9.6, and RadioScatter version 1.1.0 were used to produce results for this paper.

\section*{Acknowledgements} RET-CR is supported by the National Science Foundation under award numbers NSF/PHY-2012980 and NSF/PHY-2012989. This work is also supported by the Flemish Foundation for Scientific Research FWO-12ZD920N, the European Research Council under the EU-ropean Unions Horizon 2020 research and innovation programme (grant agreement No 805486), and the Belgian Funds for Scientific Research (FRS-FNRS). A.~Connolly acknowledges support from NSF Award \#1806923. S.~Wissel was supported by NSF CAREER Awards \#1752922 and \#2033500. DZB is grateful for support from the U.S. National Science Foundation-EPSCoR (RII Track-2 FEC, award ID 2019597). We express our gratitude to R.~Dallier, L.~Martin, J-L.~Beney and the CODALEMA experiment for providing electronics and hardware to be used in the surface radio stations of RET-CR. Computing resources were provided by the Ohio Supercomputer Center.

\bibliography{bib}

\appendix

\section{Station Layout}\label{appendix:station_layout}

\begin{figure*}[h]
  \centering
  \includegraphics[width=0.7\textwidth]{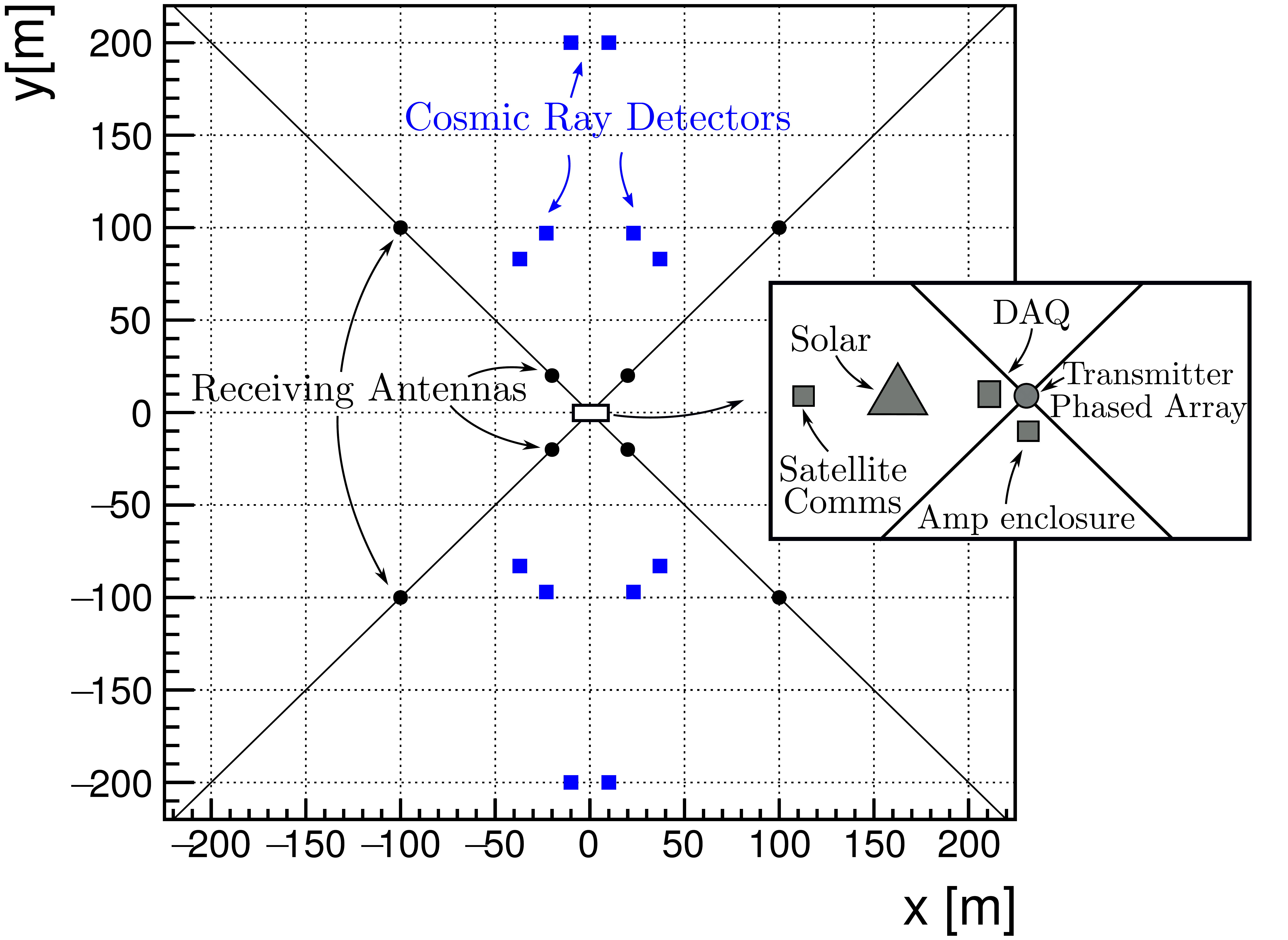}
    \caption{The station layout for RET-CR. A phased transmitter array is centrally located along with a data acquisition system (DAQ) and an amplifier enclosure for the transmitter power amplifier(s). Three 1\,kW solar power arrays oriented in a triangle are indicated along with satellite communications. The cosmic ray detector system is shown in blue. These serve to trigger the DAQ. The dimensions of the station are also indicated. }
  \label{layout}
\end{figure*}

The station layout for RET-CR is shown in Figure~\ref{layout}. A phased transmitter is centrally located along with a data acquisition system and an amplifier enclosure for the transmitter. Three 1\,kW solar power arrays are oriented in a triangle to maximize power over the full austral summer day. Satellite communications are shown near the solar power array. Each receive antenna is a vertical phased dipole array to maximize gain in an azimuthally symmetric region near the surface. The cosmic ray detector system is shown in blue, where each of the six, two-panel stations is shown in blue. The receive antennas are arranged in two sets, near at 20\,m from the TX and far at 100\,m from the TX. The drawing is not to scale.

\section{System diagram}\label{appendix:system_diagram} 

\begin{figure*}[h]
  \centering
  \includegraphics[width=0.9\textwidth]{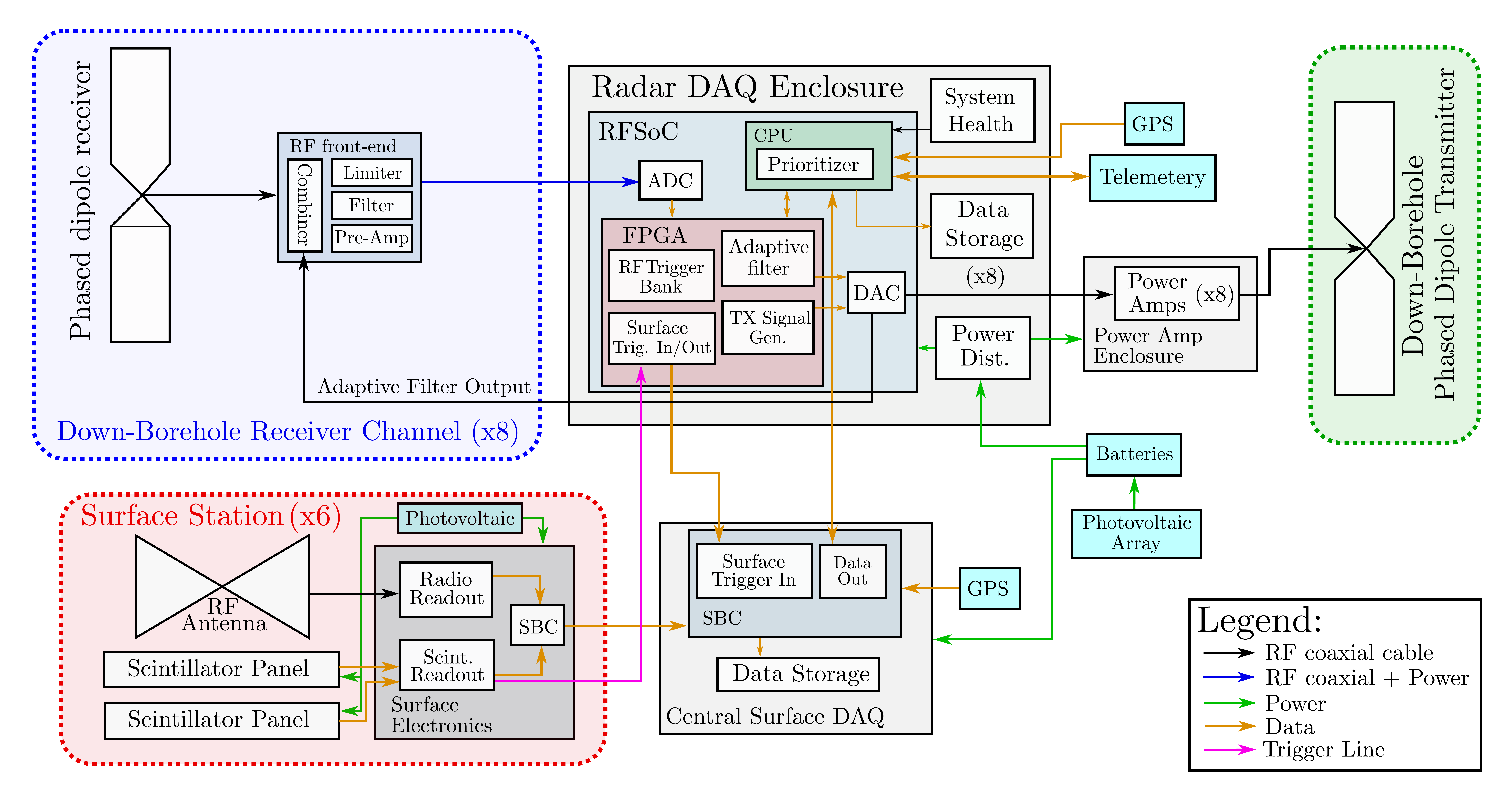}
    \caption{The system diagram for RET-CR. The down-borehole receiver system is indicated on the left. There are 8 identical receivers placed according to the layout shown in Figure~\ref{layout}. The line indicating RF cable + power going to the receiver channel is powered via the bias-tee. The down-borehole transmitter is an 8-channel phased array, each antenna having its own DAC channel and power amplifier. The surface system, shown on the bottom, has 6 identical individual stations (bottom left). }
  \label{systemDiagram}
\end{figure*}

A schematic of the RET-CR system, including the surface system, is shown in Figure~\ref{systemDiagram}.

\end{document}